\documentclass[useAMS,usenatbib]{mnras}
\usepackage{graphicx,amssymb,amsmath,times,verbatim,pdfpages,multirow,booktabs,array,dcolumn}
\usepackage[caption=false]{subfig}
\usepackage[flushleft]{threeparttable}

\title[QPO in 3C 454.3]{Multiwaveband Quasi--periodic Oscillation in the Blazar 3C 454.3}
\author[Sarkar A et al.]{Arkadipta Sarkar$^1$\thanks{sarkadipta@gmail.com}, Alok C. Gupta$^{2}$\thanks{alok@aries.res.in}, Varsha R.
  Chitnis$^{1}$\thanks{vchitnis@mailhost.tifr.res.in}, Paul J. Wiita$^{3}$ \\ 
\\
$^{1}$Department of High Energy Physics, Tata Institute of Fundamental Research, Mumbai 400005, India \\
$^{2}$Aryabhatta Research Institute of Observational Sciences (ARIES), Manora Peak, Nainital 263001, India \\
$^{3}$Department of Physics, The College of New Jersey, PO Box 7718, Ewing, NJ 08628-0718, USA}

\begin{document}
\label{firstpage}
\pagerange{\pageref{firstpage}--\pageref{lastpage}}
\maketitle

\begin{abstract}
\noindent We report the detection (\(>4\sigma\)) of a Quasi-Periodic Oscillation (QPO) in the $\gamma$-ray light curve of 3C 454.3 along with a simultaneous marginal QPO detection
(\(>2.4\sigma\)) in the optical light curves. Periodic flux modulations were detected in both of these wavebands with a dominant period of $\sim 47$ days. The \(\gamma\)-ray QPO
lasted for over 450 days (from MJD 56800 to 57250) resulting in over nine observed cycles which is among the highest number of periods ever detected in a blazar light curve. The
optical light curve was not well sampled for almost half of the \(\gamma\)-ray QPO span due to the daytime transit of the source, which could explain the lower significance of the
optical QPO. Autoregressive Integrated Moving Average (ARIMA) modelling of the light curve revealed a significant, exponentially decaying, trend in the light curve during the QPO,
along with the $47$ days periodicity. We explore several physical models to explain the origin of this transient quasi-periodic modulation and the overall trend in the observed
flux with a month-like period. These scenarios include a binary black hole system, a hotspot orbiting close to the innermost stable circular orbit of the supermassive black hole,
and precessing jets. We conclude that the most likely scenario involves a region of enhanced emission moving helically inside a curved jet. The helical motion gives rise to the QPO
and the curvature ($\sim 0.05^{\circ}$ pc$^{-1}$) of the jet is responsible for the observed trend in the light curve.
\end{abstract}


\section{Introduction}
\label{sec:1}
Blazars are the class of active galactic nuclei (AGN) that show the most substantial variability across all bands of the electromagnetic spectrum. All active galaxies are
understood to derive their ultimate power from accretion onto a supermassive black hole (SMBH) with a mass in the range of 10$^{6}$ -- 10$^{10}$ M$_{\odot}$. Blazars also possess
relativistic jets pointing toward us that dominate the observed emission in most bands \citep{UrryPadovani1995, 1995ARA&A..33..163W, 2019ARA&A..57..467B}. There are quite a few
similarities in the nature of the time series data (light curves) between X-ray emission from AGN and X-ray emitting binaries in our and nearby galaxies, where gas flows from a
star through an accretion disc onto a neutron star or black hole of several solar masses. In fact, AGNs can be considered as scaled-up galactic BH binaries wherein the location of
a break from shallower to steeper slopes in the red-noise power-spectrum of the light curve is proportional to the BH mass \citep{2004ApJ...609L..63A}. Although QPOs are rather
common in the light curves of X--ray emitting binaries \citep{2006ARA&A..44...49R}, they appear to be quite rare for AGNs \citep[see][for reviews]{2014JApA...35..307G,
2018Galax...6....1G}.

\begin{figure*}
  \centering
  \includegraphics[width=.9\linewidth]{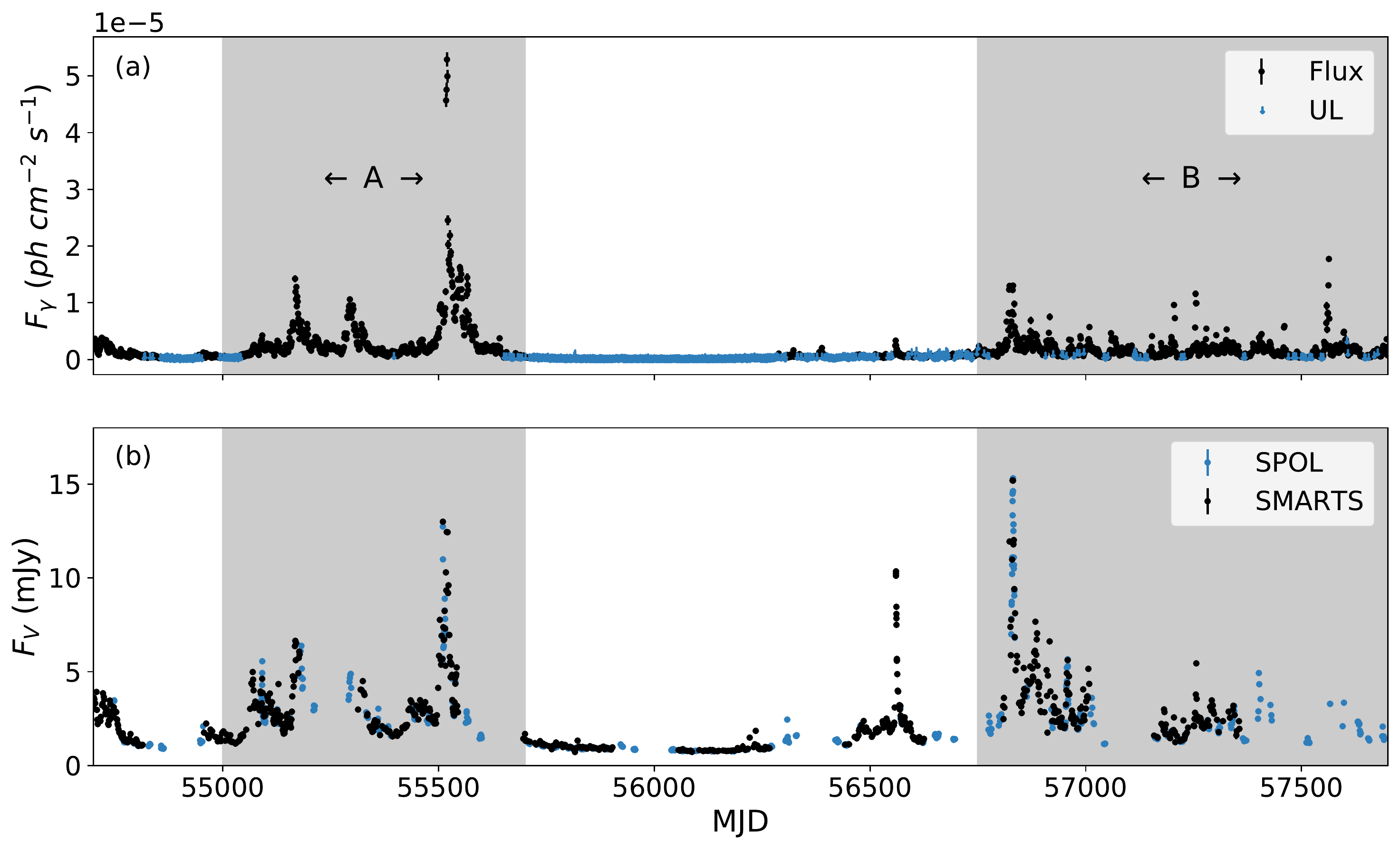}
  \caption{\label{f1}
    \textbf{(a)} The 0.1--300 GeV $\gamma$-ray light curve of the blazar 3C 454.3 from \textit{Fermi}--LAT in 1--day bins. The blue points are the upper limit values where the
detection significance is $< 5\sigma$. Searches for QPOs were performed in the shaded intervals A and B during which the source was almost always detected in $\gamma$-rays.
\textbf{(b)} Optical V band light curve for this source from SMARTS (black) and SPOL (blue). The y-axis is in milli-Jansky ($10^{-26}$\ erg\ cm$^{-2}$\ s$^{-1}$\ Hz$^{-1}$). It is
clear that when independent optical measurements were taken on the same nights there is excellent agreement between them. Search for a QPO revealed significant periodicity in
segment B.}
\end{figure*}

There have been rather strong claims of AGN QPOs in different bands of the electromagnetic spectrum, ranging from minutes through days through months and years \citep[e.g.,][and
references therein]{2008Natur.455..369G, 2009A&A...506L..17L, 2009ApJ...690..216G, 2018A&A...616L...6G, 2019MNRAS.484.5785G, 2013MNRAS.436L.114K, 2014JApA...35..307G,
2018Galax...6....1G, 2015ApJ...813L..41A, 2016ApJ...819L..19P, 2018NatCo...9.4599Z, 2019MNRAS.487.3990B}. However, many of the claimed QPOs, particularly those made earlier, were
marginal detections \citep{2014JApA...35..307G}, lasting only a few cycles, and the originally quoted statistical significances are probably overestimates
\citep{2014JApA...35..307G, 2019MNRAS.482.1270C}. Among the better recent claims of QPOs in the $\gamma$-ray band are of $\sim$34.5 days in the blazar PKS 2247--131
\citep{2018NatCo...9.4599Z} and of $\sim$ 71 days in the blazar B2 1520+31 \citep{2019MNRAS.484.5785G} found as part of a continuing analysis of blazar \textit{Fermi}--LAT
observations. A recent claim of a $\sim$44 day optical band QPO in the narrow-line Seyfert 1 galaxy KIC 9650712 from densely sampled {\it Kepler} data has been made by
\citet{2018ApJ...860L..10S}; it was supported by an independent analysis of the same data, indicating a QPO contribution at 52$\pm$2 days \citep{2020arXiv200102800P}. Some possibly
related QPOs of a few hundred days in two widely separated bands have been reported \citep{2016AJ....151...54S, 2016ApJ...820...20S, 2017A&A...600A.132S}. However, an analysis of
the \textit{Fermi}--LAT and aperture photometry light curves by \citet{2019MNRAS.482.1270C} argued that some multiwaveband QPOs, along with many earlier claims of $\gamma$-ray
QPOs, are not significant. Among the \(\gamma\)-ray QPOs with month-like periods, none showed simultaneous oscillations in a different wavebands. Evidence for related QPOs in
multiple wavebands was observed in PG 1553\(+\)113, where a QPO was detected in the 0.1--300 GeV and the optical waveband \citep{2015ApJ...813L..41A}. The observed QPO had a
dominant period of \(\sim 754\) days and the source showed strong inter-waveband cross correlations.

3C 454.3 (B2251+158) is among the brightest and most well studied of the flat spectrum radio quasar (FSRQ) subclass of blazars and is situated at a redshift of $z =$ 0.859
\citep{1989ApJS...69....1H}. The mass of the SMBH at the center of 3C 454.3 is estimated by optical spectroscopy methods to be in the range of (0.5 -- 2.3) $\times$ 10$^{9}$
M$_{\odot}$ \citep{2017MNRAS.472..788G, 2019A&A...631A...4N}. The flow speed down the approaching relativistic jet is in the range of 0.97c to 0.999c \citep{2005AJ....130.1418J,
2009A&A...494..527H} and the angle to the observer's line of sight is between 1$^{\circ}$ and 6$^{\circ}$ \citep{2019ApJ...887..185S}. 3C 454.3 has shown a variety of observational
patterns in multiband observations made at different epochs, including correlated multiwaveband emission (in all wavebands except X-rays) \citep{2009ApJ...697L..81B,
2012AJ....143...23G, 2017MNRAS.464.2046K} with a strong $\gamma$-ray flaring event. It was modeled using a combination of standing conical shocks and magnetic reconnection events
in the core of the jet \citep{Jorstadetal2013}. A strong, correlated multiwaveband flare occurred in 2009 December 3--12 where dramatic changes ($\sim 170^{\circ}$) in optical
polarization angle were observed along with strong anti-correlation between optical flux and degree of polarization \citep{2017MNRAS.472..788G}. It is a peculiar blazar, and
various sets of possible correlations between multiwavelength observations at different epochs have been noted \citep[e.g.,][and references therein]{2017MNRAS.472..788G,
2019ApJ...887..185S, 2019A&A...631A...4N}.

Here we detect in 3C 454.3, for the first time, a \(\sim 4\sigma\) QPO, with a month-like period of around 47 days, simultaneously in both archival $\gamma$-ray measurements and
optical V-band observations. We examine several possible explanations for this periodicity and conclude that a geometrical model with blobs moving helically in a curved jet best
explains this QPO aspect of the variability, and where the curvature of the relativistic jet of 3C 454.3 is estimated from the flux modulation during the QPO. The subsequent
sections of this work deal with the methodology involved in the detection of the QPO and its physical interpretation. The data reduction and QPO detection methodology are given in
Section \ref{sec:2} and Section \ref{sec:3}, respectively. The results are elaborated in Section \ref{sec:4} with our physical interpretation given in Section \ref{sec:5}.

\section{Data Acquisition}
\label{sec:2}

The $\gamma$-ray observations were made using the Large Area Telescope onboard the \textit{Fermi} satellite (\textit{Fermi}--LAT). It is a pair conversion detector having a field
of view of 2.4 sr \citep{2009ApJ...697.1071A} and provides near constant monitoring of the $\gamma$-ray sky. For our analysis, we used the \texttt{Fermitools
(v1.2.23)}\footnote{\href{https://github.com/fermi-lat/Fermitools-conda/}{https://github.com/fermi-lat/Fermitools-conda/}} package, selecting events within the energy range 100 MeV
to 300 GeV from a circular Region of Interest (ROI) with radius $15^{\circ}$ centered around the source (RA: 343.491, Dec: $+$16.1482). We filtered the data using the criterion
\texttt{(DAT\_QUAL $> 0$ \&\& LAT\_CONFIG == 1)} and employed a zenith angle cut of \(90^{\circ}\) to prevent source contamination from the earth limb. The data were fitted by an
unbinned likelihood method using the tool \texttt{gtlike} \citep{1979ApJ...228..939C, 1996ApJ...461..396M} which gave the significance of each source within the ROI in the form of
Test Statistics \citep[TS \(\sim \sigma^2\),][]{1996ApJ...461..396M}. During light curve generation, data in each time bin was iteratively fitted and the sources with low
significance (TS \(<\) 1) were rejected prior to the next iteration until the fit converged. The light curve was obtained by integrating the source fluxes with an integration time
of 1 day for the intervals where the TS exceeded 25 (over $\sim 5 \sigma$ significance). For the light curve generation, the source spectrum was modeled using a log-parabola, whose
index was kept free during the fit. The spectral parameters of the 188 point sources in the ROI were taken from the 4FGL catalog. During fitting, the spectral parameters of dim (TS
\(<\) 25) and relatively distant sources (beyond \(5^{\circ}\)) were kept fixed, with the exception of the bright source J2232.4+1143 (CTA 102) at about \(12^{\circ}\) from 3C
454.3. We modeled the galactic diffuse emission using \texttt{gll\_iem\_v07.fit} and the isotropic background using \texttt{iso\_P8R2\_SOURCE\_V6\_v06.txt}. The iterative fitting
of light curves was carried out using the \texttt{enrico}\footnote{\href{https://github.com/gammapy/enrico}{https://github.com/gammapy/enrico}} software
\citep{2013arXiv1307.4534S}.

Because of the relative brightness of the blazar 3C 454.3, several long-term ground-based optical monitoring programs have tracked its variability for many years, albeit with
unavoidable annual gaps when it is not visible at night. We obtained magnitudes of 3C 454.3 for the same interval as the \textit{Fermi} data were available from the public archive
of the Small and Medium Aperture Research Telescope System (SMARTS). These observations were taken at the Cerro Tololo Inter-American Observatory in Chile using CCD imaging and
photometry on the 1.3-m telescope. Details about the instrument, observation, and data reduction and analysis of SMARTS data are described in \cite{2012ApJ...756...13B}. We also
used optical photometric observations that were carried out at Steward Observatory, University of Arizona, using SPOL (a CCD Imaging/Spectropolarimeter). Details of this
instrument, observations, and data analysis are provided in \cite{2009arXiv0912.3621S}. The V magnitudes of 3C 454.3 obtained from SMARTS and Steward Observatory observations were
combined into 1-day bins after converting the magnitude values into flux densities.

The 3C 454.3 light curves are given in Figures \ref{f1}a am=nd b. The light curves were divided into segments A (MJD 55000 to 55700) and B (MJD 56750 to 57700), periods during which
the $\gamma$-ray emission was high enough to usually allow for strong detections on individual days. A visual inspection of Figure \ref{f1} indicated the possible presence of a
quasi-periodic flux modulation during segment B in both $\gamma$-ray and optical wavebands. We performed several quantitative tests to verify this indication of periodicity.

\begin{figure*}
  \centering
  \includegraphics[width=0.9\linewidth]{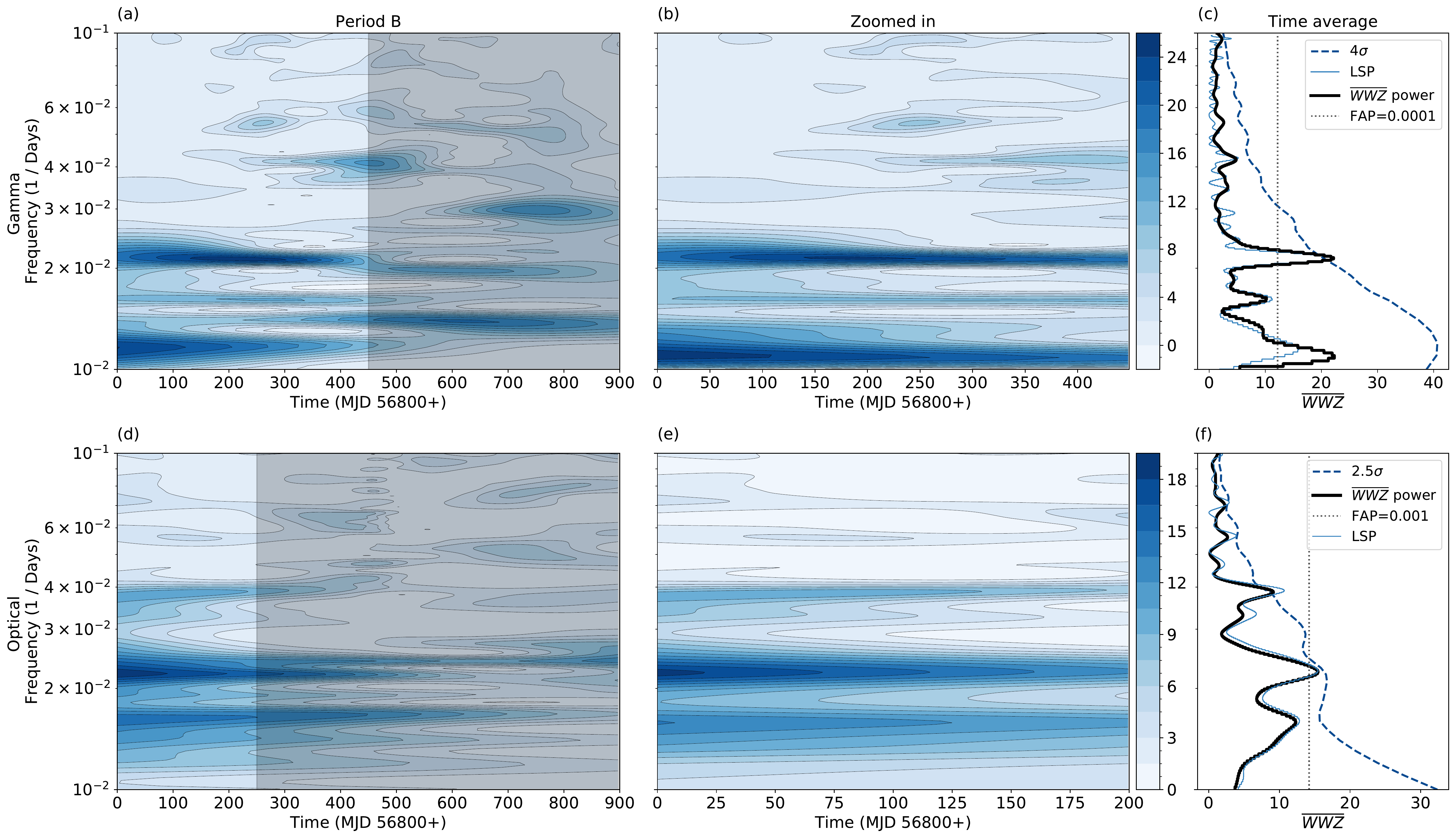}
  \caption{\label{f2} Wavelet analyses for the $\gamma$-ray (top row) and optical (bottom row) light curves of 3C 454.3. \textbf{(a)} WWZ map for the entire interval B for the
$\gamma$-rays showing a strong signal around 47 d over the first half of the data (450 d), which is the unshaded region. \textbf{(b)} Zoomed in WWZ map for the unshaded region in
\textbf{(a)}, showing a strong signal is present for over 9 cycles. \textbf{(c)} Time averaged WWZ and LSP for the region in \textbf{(b)}, yielding a strong signal of a
$\gamma$-ray QPO of $\sim 4.1 \sigma$ significance. \textbf{(d)} WWZ map for the entire interval B for the V-band emission showing a signal around 47 d over the first 250 days of
the observations (unshaded region). \textbf{(e)} Zoomed in WWZ map for the unshaded region in \textbf{(d)}, showing a strong signal is present for over 4 cycles. \textbf{(f)} Time
averaged WWZ and LSP for the region in \textbf{(e)}, yielding a good indication of a QPO of identical period in the optical, with $\sim 2.4\sigma$ significance.}
\end{figure*}

\begin{figure*}
  \centering
  \includegraphics[width=0.9\linewidth]{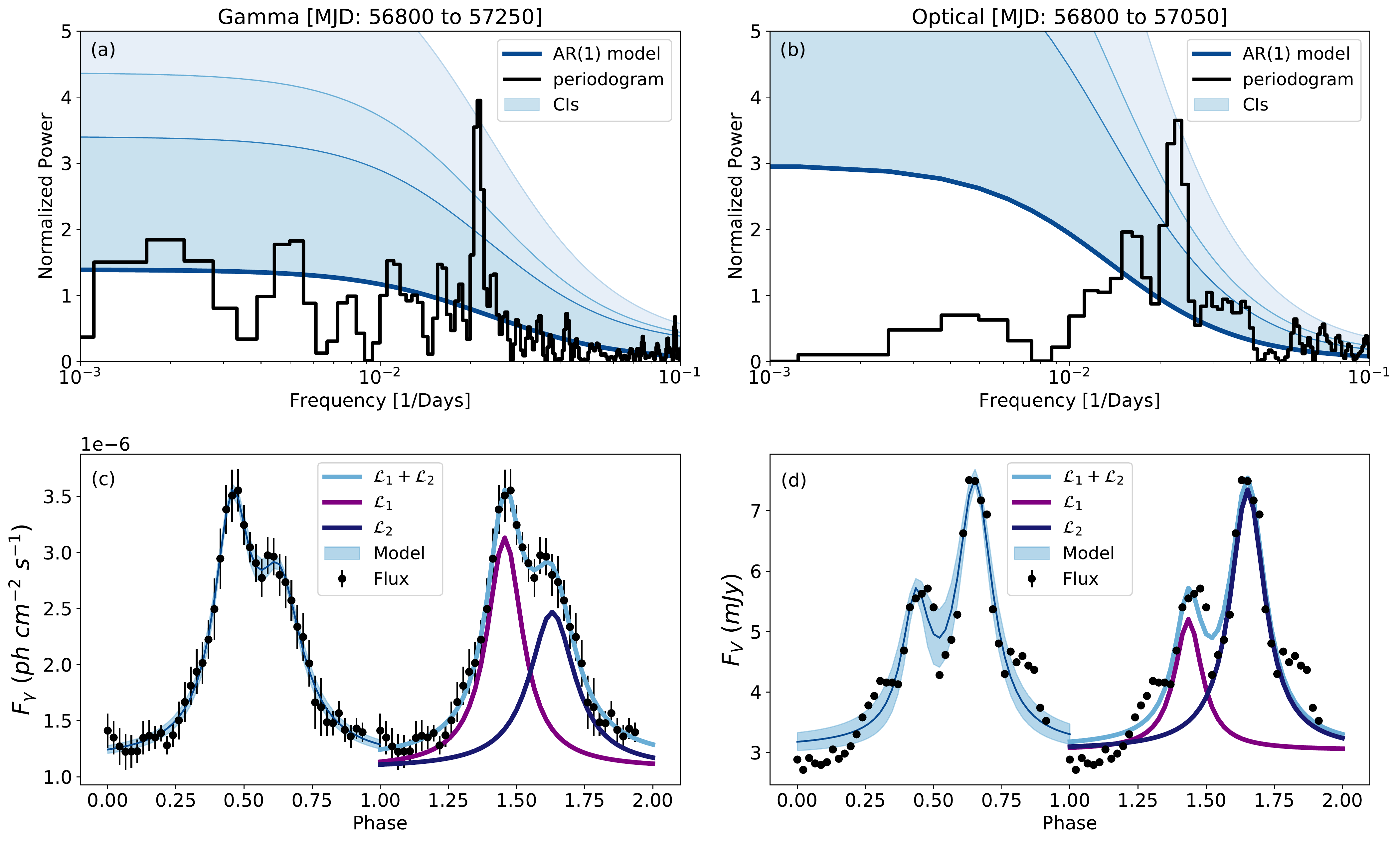}
  \caption{\label{f3} \textbf{(top)} Lomb--Scargle periodograms (black) for \textbf{(a)} $\gamma$-ray and \textbf{(b)} optical V--band light curves. The deep blue curves give the
theorteical AR(1) models (Equation \ref{eq4}) and the blue shaded regions are the $\chi^2$ confidence interval about the theoretical model corresponding to FAPs of $0.1$, $0.05$
and $0.01$ respectively (darker to lighter). \textbf{(bottom)} Folded light curves (black) \textbf{(c)} for $\gamma$-rays and \textbf{(d)} for optical emission fitted using models
(light blue) with two Lorentzian components (purple and dark blue). The time ranges considered are at the tops of the columns.}
\end{figure*}

\begin{figure*}
  \centering
  \includegraphics[width=0.9\linewidth]{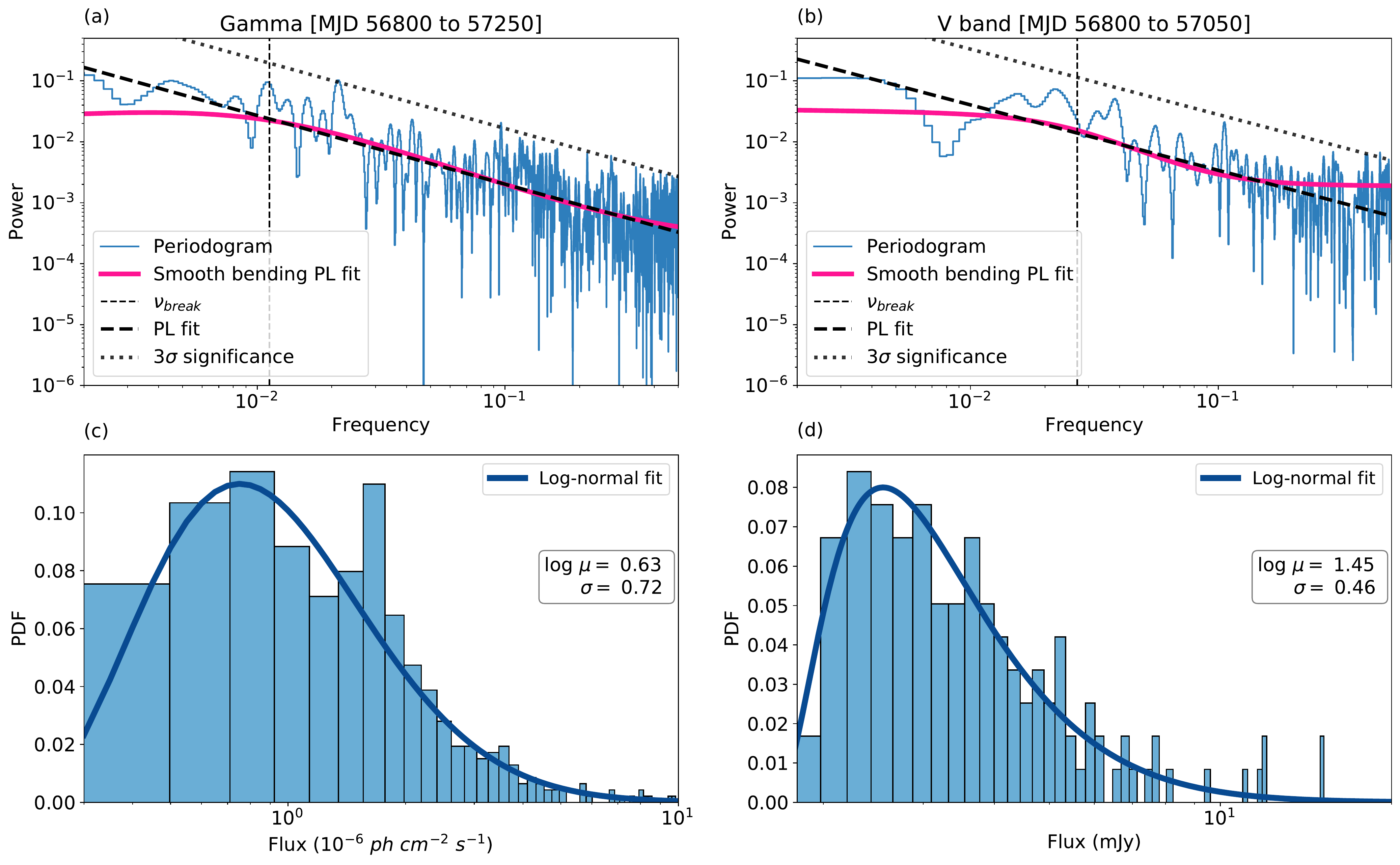}
  \caption{\label{f9} \textbf{(a)} PSD of the $\gamma$-ray light curve during the QPO. The PSD (blue) is fitted with a power-law (black dashed) and a smooth bending power-law model
(red), with a break at $\nu_{break}$ (vertical dashed line). The dotted line is the $3\sigma$ significance threshold (for the dominant period) calculated analytically considering
an underlying power-law model following \citet{2005A&A...431..391V} \textbf{(b)} Same as \textbf{(a)} but for the V-band. \textbf{(c)} $\gamma$-ray flux distribution during QPO
fitted with a lognormal model (dark blue), with parameters inset in the panel. \textbf{(d)} Same as \textbf{(c)} for the V band. The parameters of the fitted PSD and PDF were used
to simulate light curves.}
\end{figure*}

\begin{figure*}
  \centering
  \includegraphics[width=0.9\linewidth]{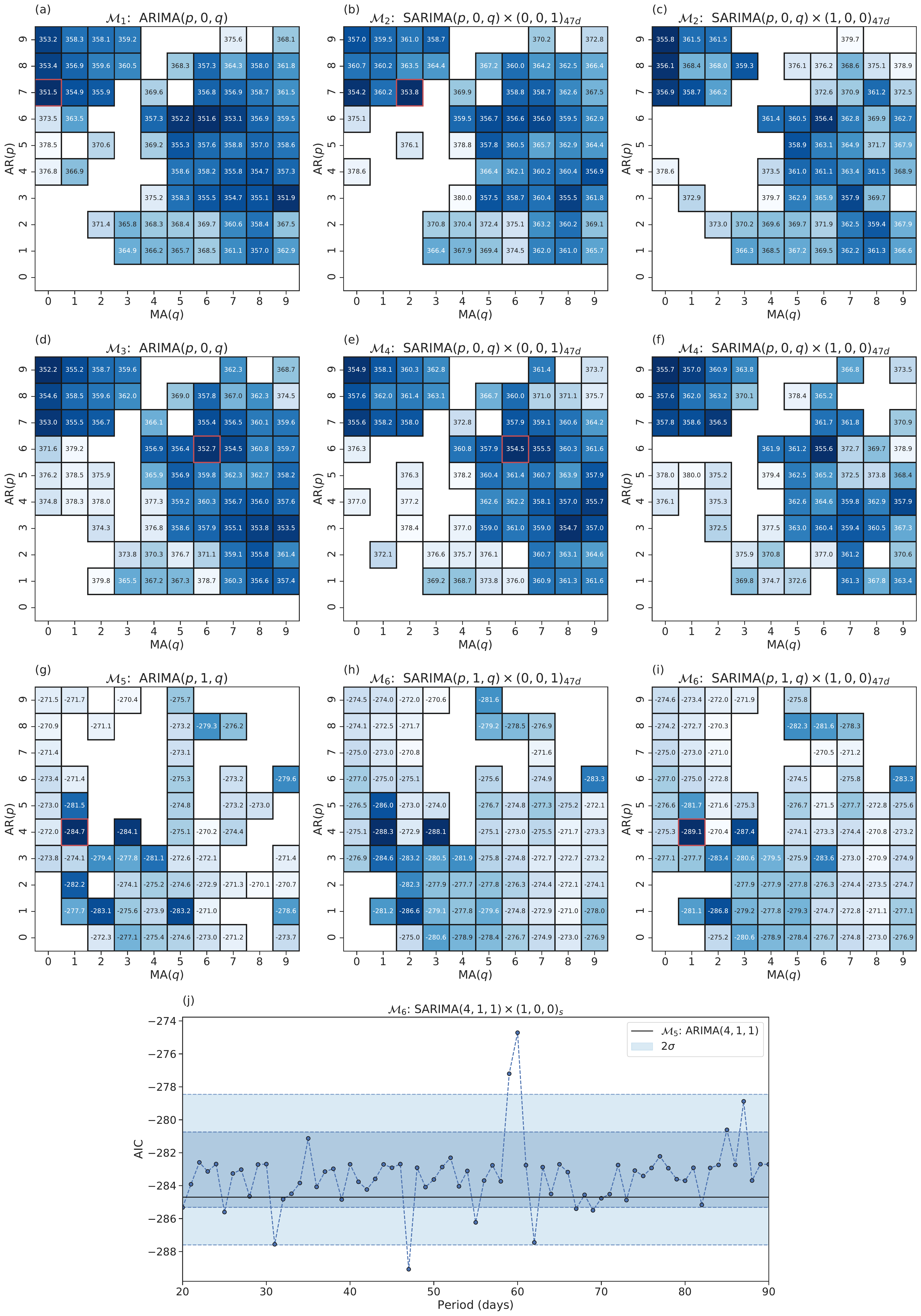}
  \caption{\label{f4} \textbf{(a)}--\textbf{(i)} AIC maps for a subset of models in the classes $\mathcal{M}_{1-6}$, with darker shades corresponding to better models. The best
model in each class is bordered in red and we observe the best AIC value to arise from the model $\mathcal{M}_6$: SARIMA$(4,1,1)\times(1,0,0)_{47\text{d}}$. \textbf{(j)} AIC values
for a subset of models $\mathcal{M}_6$: SARIMA$(4,1,1)\times(1,0,0)_{s}$ with different periodicities ($s$). We see a significant dip in the AIC value at $s=47$ days. The
horizontal black line corresponds to the AIC value for the non periodic control model $\mathcal{M}_5$: ARIMA$(4,1,1)$ and the shaded egions denote the $1\sigma$ and $2\sigma$
extents. All the AIC values are shifted following, AIC$_{min} + $AIC$_{max}/2 \approx 100$, to improve visibility.}
\end{figure*}

\begin{figure*}
  \centering
  \includegraphics[width=0.9\linewidth]{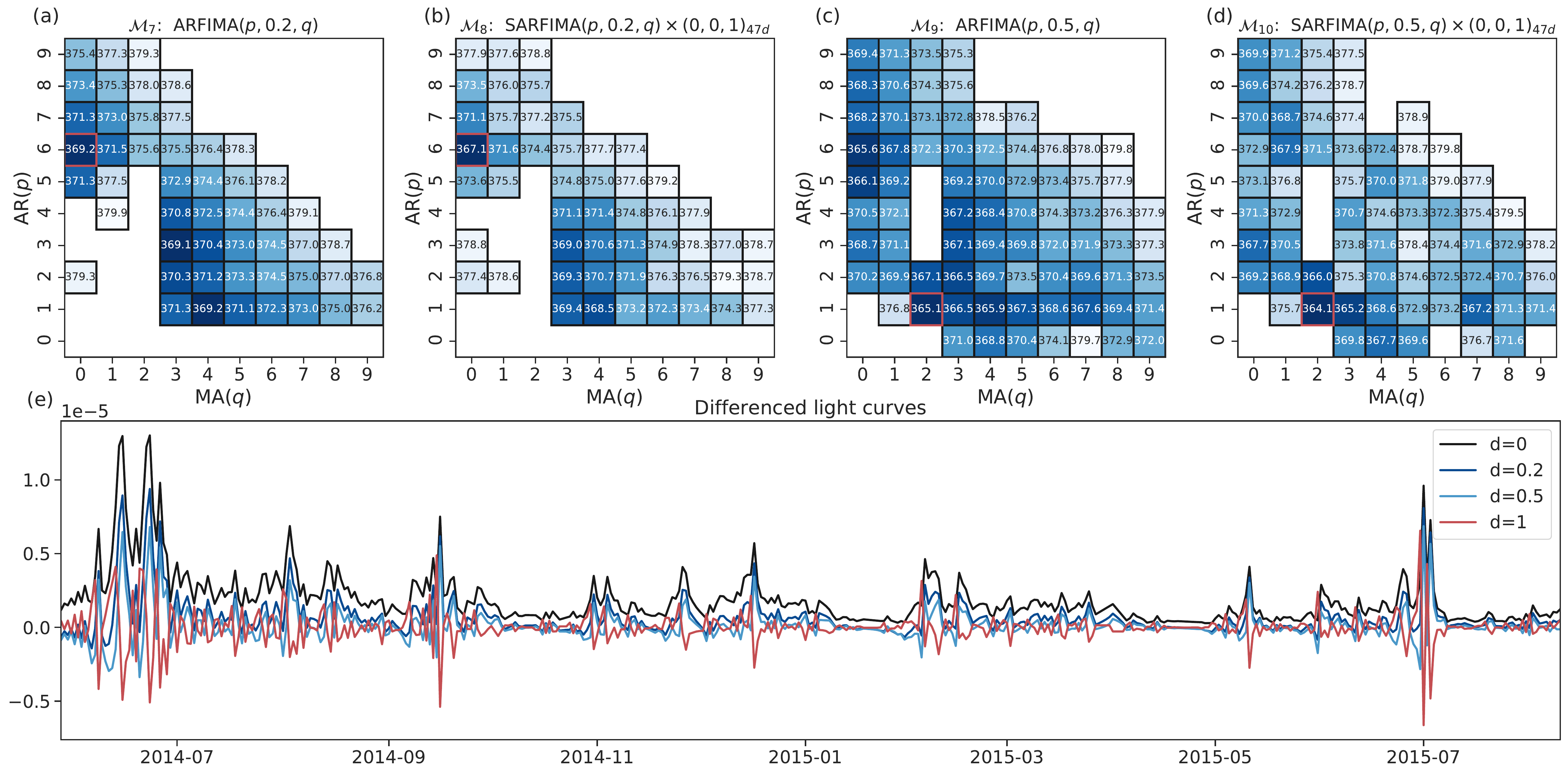}
  \caption{\label{f5} \textbf{(a)}--\textbf{(d)} AIC maps for a subset of models in the classes $\mathcal{M}_{7-10}$. The AIC values are shifted globally as mentioned in the
Figure \ref{f4} caption to improve readability. \textbf{(e)} Differenced light curve with different orders of differencing used to fit an SARIMA$(p,0,q)$ process, which is
equivalent to fitting a SARFIMA model to the light curve.}
\end{figure*}

\begin{figure*}
  \centering
  \includegraphics[width=0.9\linewidth]{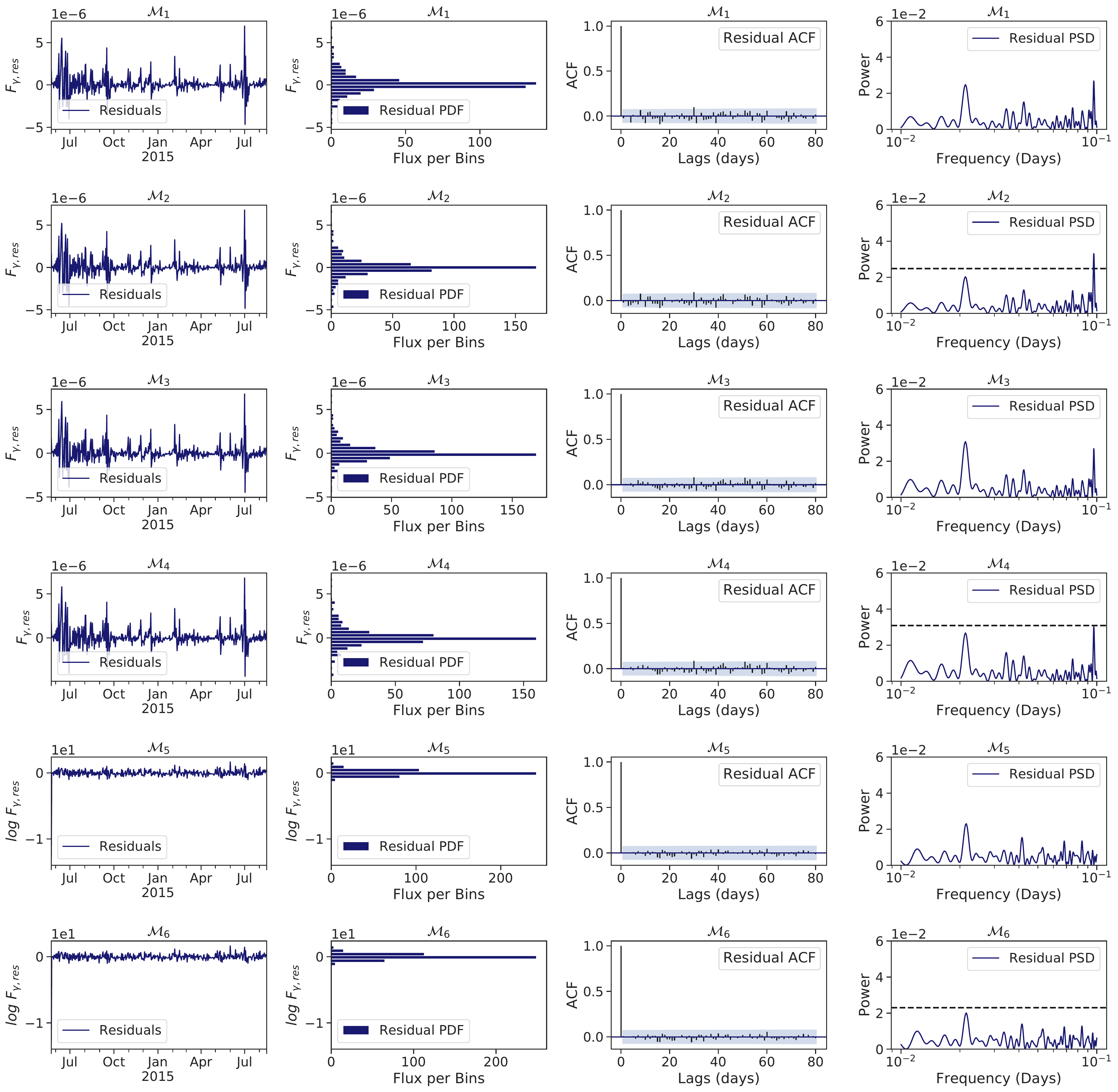}
  \caption{\label{f6} (left to right) Residual, PDF, ACF, and PSD of the residuals obtained from fitting the $\gamma$-ray light curve using the best models from each class
($\mathcal{M}_{1-6}$). The class increases from top to bottom. We see that periodic model residuals have lower PSD peak power at $\sim 47$ days compared to the corresponding
nonperodic models (black horizontal dashed lined).}
\end{figure*}

\begin{figure*}
  \centering
  \includegraphics[width=.9\linewidth]{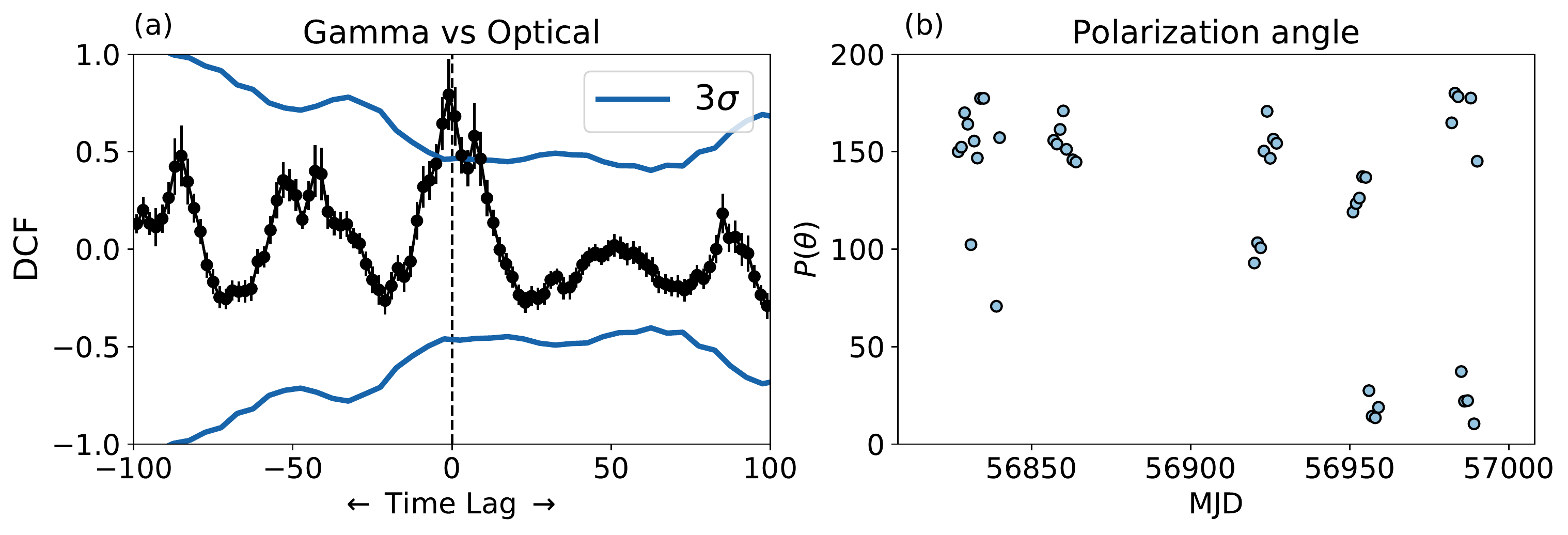}
  \caption{\label{f1x}
    \textbf{(a)} $\gamma$-ray and V-band emission DCF during segment B. \textbf{(b)} Optical polarization angle during the detected QPO interval.}
\end{figure*}

\section{Analysis}
\label{sec:3}

One of the most common tests to detect periodicities in astrophysical light curves is the periodogram, which gives the power spectral density (PSD) or the power emitted in
different frequencies. For a regularly sampled time series, the periodogram is the mod-square of its discrete Fourier transform (DFT). For an irregularly sampled time series, as is
the case with most astrophysical light curves, the periodogram is obtained by iteratively fitting sinusoids with different frequencies to the light curve and constructing a
periodogram from the goodness of the fit. This technique is known as the Lomb-Scargle Periodogram \citep[LSP,][]{1976Ap&SS..39..447L, 1982ApJ...263..835S}. Periodograms are
effective tools in measuring the presence of persistent periodicities. However, they often fail to detect transient periodicities if the time range of periodic modulation is not
identified \textit{a priori} since the non-periodic span decreases the LSP goodness of fit, thereby decreasing the power of the dominant period.

For identifying transient periodicities, we used a wavelet method, Weighted Wavelet Z-transform (WWZ) to decompose the data into time and frequency domains \citep[WWZ maps,
][]{1996AJ....112.1709F}, by convolving the light curve with a time and frequency-dependent kernel. We used the abbreviated Morlet kernel \citep{doi:10.1137/0515056} which has the
following functional form: $f(\omega[t-\tau])=\exp\left[i \omega(t-\tau)-c \omega^{2}(t-\tau)^{2}\right]$. The WWZ map is then given by
\begin{equation}
  \label{eq1}
  W(\omega, \tau ; x(t)) =\omega^{1 / 2} \int x(t) f^{*}(\omega(t-\tau)) d t .
\end{equation}
Here $f^*$ is the complex conjugate of the wavelet kernel $f$, $\omega$ is the scale factor (frequency) and $\tau$ is the time-shift. This kernel is like a windowed DFT with
the window $\exp\left[-c \omega^{2}(t-\tau)^{2}\right]$, where the window size depends on $\omega$ and the constant $c$. A WWZ map has the advantage of locating both any dominant
periods and their spans in time.

To quantify the significance of the dominant PSD peak, it is necessary to assume an underlying model for the periodogram that gives rise to the observed PSD. The significance of
the dominant period can then be estimated from the probability of obtaining the dominant peak power from statistical fluctuations about the underlying PSD model. Futhermore, since
different light curves will have different PSDs, a stochastic model for the light curve is also necessary to comment on the underlying PSD. Time series, such as blazar light
curves, are often modeled by an Autoregressive (AR) process, where the present emission is related to the past emissions, thereby making large sudden fluctuations in emission less
likely \citep{ROBINSON19779}.

Considering \(\mathcal{F}(t_i)\) to be the emission at time \(t_i\) and \(\epsilon(t_i)\) a normally distributed random variable representing the fluctuations (or fit errors), the
AR light curves are analytically given by: \(\mathcal{F}(t_i) = \sum_{j=1}^p \theta_j \mathcal{F}(t_{i-j}) + \epsilon(t_i)\), where \(p\) is the order of the AR process considered,
denoting the time lag over which the past emissions affect the present emission and \(\theta_j\)'s are the AR coefficients. As previously mentioned, for an AR process, large sudden
fluctuations are less likely. Thus, we expect higher powers to be present at lower frequencies, giving rise to an underlying red-noise PSD. Similarly, we can also have a Moving
Average (MA) model for the light curve where the present emission depends on the past fluctuations. It can be given by an analogous expression: \(\mathcal{F}(t_i) = \sum_{j=1}^q
\phi_j\epsilon(t_{i-j}) + \epsilon(t_i)\) where \(q\) is the order of the MA and the \(\phi_j\)'s are the MA coefficients.

The AR and MA models can be combined to form an Autoregressive Moving Average (ARMA) model which can explain quite general stationary time series. It is best represented by
considering a lag operator (\(\mathcal{L}\)) given by: \(\mathcal{L}^k\mathcal{F}(t_i)=\mathcal{F}(t_{i-k})\). Emperically, blazar light curves are non-stationary, and to deal with
this, one then can perform successive differencing (\(\Delta\)) of order \(d\), \(\Delta^d\mathcal{F}(t_i) = (1-\mathcal{L})^d\mathcal{F}(t_i)\), prior to fitting the light curve
to the ARMA model. This process of differencing the light curve is semantically equivalent to Integrating (I) the model, resulting in an Autoregressive (AR) Integrated (I) Moving
Average (MA) or ARIMA\((p,d,q)\) model \citep[][]{1981ApJS...45....1S, 2018FrP.....6...80F}. The analytical representation of the ARIMA\((p,d,q)\) model is then:
\begin{equation}
  \label{eq2}
  \begin{split}
    \left(1 - \sum_{j=1}^p \theta_j \mathcal{L}^j \right)\Delta^d \mathcal{F}(t_i) = \left(1 - \sum_{j=1}^q \phi_j \mathcal{L}^j \right) \epsilon(t_i),
  \end{split}
\end{equation}
where the first term on the left hand side (LHS) of Equation \ref{eq2} is due to the AR process, the term with \(\Delta\) is the successive differencing, and the right hand side
(RHS) is due to the MA process.

\begin{table*}
    \caption{\label{t1} Best fitting model from each class}
    \resizebox{\textwidth}{!}{
      \begin{threeparttable}
  \begin{tabular}{|c|cc|cc|cc|cccc|}
    \hline
Class: & $\mathcal{M}_1$ & $\mathcal{M}_2$ & $\mathcal{M}_3$ & $\mathcal{M}_4$ & $\mathcal{M}_5$ & $\mathcal{M}_6$ & $\mathcal{M}_7$ & $\mathcal{M}_8$ & $\mathcal{M}_9$ & $\mathcal{M}_{10}$\\
  Trend:  & \multicolumn{2}{c|}{No Trend} & \multicolumn{2}{c|}{Additive} & \multicolumn{2}{c|}{Multiplicative} & \multicolumn{4}{c|}{Long Term Memory}   \\
    \hline
      BFO:    & $(7,0,0)$  & $(7,0,2)$  & $(6,0,6)$  & $(6,0,6)$ & $(4,1,1)$ & $(4,1,1)$ & $(6,0.2,0)$ & $(6,0.2,0)$ & $(1,0.5,2)$ & $(1,0.5,2)$ \\
      BFSO:   & -          & $(0,0,1)$  & -          & $(0,0,1)$ & -         & $(1,0,0)$ & -           & $(0,0,1)$   & -           & $(0,0,1)$   \\
      Period: & -          & $47$d      & -          & $47$d     & -         & $47$d     & -           & $47$d       & -           & $47$d       \\
      \hline
      AIC:    & $351.5$    & $353.8$    & $352.7$    & $354.5$   & $-284.7$  & $-289.1$  & $369.2$     & $367.1$     & $365.1$     & $364.1$     \\
      \hline
  \end{tabular}
  \begin{tablenotes}
      \small
      \item \texttt{NOTE:} BFO: best fitting order $(p,d,q)$ of an ARIMA model. BFSO: best fitting seasonal order $(P,D,Q)$ of the SARIMA model. Additive trends are
        linear, $A(t)=mt+c$, and multiplicative trends are exponential, $\log A(t) = mt+c$. Fractionally integrated models were considered to incorporate the longer term memory
        models. The AIC values are shifted as mentioned in the Figure \ref{f4} caption.
    \end{tablenotes}
  \end{threeparttable}
}
\end{table*}

The analytical form of an ARIMA$(p,d,q)$ PSD is not available except for the simplest of cases. Considering a white-noise, or ARIMA$(0,0,0)$ process, where
$\mathcal{F}(t_i)=\phi \epsilon(t_i)$, the obtained periodogram powers are independent and normally distributed at different frequencies. Considering $N$ frequencies, the
probability ($p$) of the maximum power crossing a set threshold ($z$) was calculated using $p(>z) \approx N\ e^{-z}$ which is the False Alarm Probability (FAP) for the
threshold considered \citep{1982ApJ...263..835S, 2018AJ....155...31H}. Empirically, blazar PSDs (like most ARIMA process PSDs) fundamentally follow a red-noise process, with more
power being emitted at lower temporal frequencies, and hence the FAP due to a white-noise model could overestimate the significance of the dominant period. Another case where the
analytical form of the PSD is known is for an AR1 or ARIMA$(1,0,0)$ process. It is given by \citep{percival_walden_1993}
\begin{equation}
  \label{eq4}
  G_{rr}(f_j)=G_0\frac{1-\theta^2}{1-2\theta \cos (\pi f_i/f_{Nyq}) + \theta^2},
\end{equation}
where $f_j$ is the discrete frequency up to the Nyquist frequency ($f_{Nyq}$), $G_0$ is the average spectral amplitude, and $\theta\equiv \exp[\langle t_i-t_{i-1}\rangle/\tau]$ is
the autoregression coefficient, obtained by averaging over the sampling interval, while $\tau$ comes from Welch-overlapped-segment-averaging \citep[WOSA,][]{1161901} of the LSP.
Equation \ref{eq4} is a better model for blazar PSDs (compared to a white-noise model) since it represents a red-noise process. Considering this underlying model, the significance
of the dominant PSD peak was obtained from the $\chi^2$ distribution confidence intervals (CI) about the theoretical PSD \citep[using the software REDFIT,
\footnote{\href{https://www.manfredmudelsee.com/soft/redfit/index.htm}{https://www.manfredmudelsee.com/soft/redfit/index.htm}}][]{Schulz:2002:RER:607225.607238}. Another viable
method to estimate the significance is the Monte-Carlo (MC) method where a thousand light curves were simulated following the PSD and flux distribution (PDF) of the original light
curve \citep{2013MNRAS.433..907E}. The significance of the dominant period was estimated from the mean and standard deviation of the distribution of the simulated light curve PSD
at each frequency. It gave the odds of obtaining the observed dominant period power considering the underlying PSD model.

An independent method to identify periodicity in a light curve is to fit the light curve using a periodic extension of the ARIMA family of models,  Seasonal (S) Autoregressive
(AR) Integrated (I) Moving Average (MA) models or SARIMA$(p,d,q)\times(P,D,Q)_s$ \citep[][]{Adhikari2013AnIS, 6676239}. These are mathematically defined as
\begin{equation}
  \label{eq5}
  \begin{split}
  &\left( 1 - \sum_{j=1}^p \theta_j \mathcal{L}^j \right) \left( 1 - \sum_{j=1}^{P} \Theta_j \mathcal{L}^{sj} \right) \Delta^d\Delta_s^D\mathcal{F}(t_i) \\
   = &\left( 1 + \sum_{j=1}^p \phi_j \mathcal{L}^j \right) \left( 1 + \sum_{j=1}^{Q} \Phi_j \mathcal{L}^{sj} \right) \epsilon(t_i) + A(t),
  \end{split}
\end{equation}
where the second factors in the LHS and RHS are the seasonal AR and MA terms (with order $P$ and $Q$) responsible for periodicity with a period $s$. Here, $\Theta$ and $\Phi$ are
the seasonal AR and MA coefficients respectively and $\Delta^D_s$ is the seasonal differencing operator given by: $\Delta_s^D\mathcal{F}(t_i) =
(1-\mathcal{L}^s)^{D}\mathcal{F}(t_i)$.

Models with periodic contributions such as SARIMA$(p,d,q)\times(P,D,Q)_s$ are a superset of ARIMA$(p,d,q)$ models where the present emission further depends on the emissions or
fluctuations from the same phase in earlier periods. While searching for QPOs, the goodness of fit of a periodic model (SARIMA) was compared to a non-periodic control model
(ARIMA). The models were quantitatively compared using the Akaike information criterion \citep[AIC, ][]{1100705} defined as AIC$=-2lnL+2k$, with $L$ being the likelihood of
obtaining the light curve given the model and $k$ is the number of free parameters in the model. AIC penalizes a model for using a larger number of parameters and rewards it for
fitting the data better. Hence, during the comparison, the model with the lower AIC is favored. We performed a grid search of AIC values for all models in the parameter space
\begin{equation}
  \label{eq6}
\Psi=
    \begin{cases}
      p,q &\in [0,10]\\
      P,Q &\in [0,6]\\
      d,D &\in \{0,1\}\\
      s &\in [0,100]\ \text{days} .      
    \end{cases}
  \end{equation}

Along with the processes mentioned above, light curves can have overall trends $A(t)$. Eqn. (\ref{eq5}) is an example with an additive trend since it can be represented as
$\mathcal{F}(t_i) = R + S + A(t_i)$ by grouping all the seasonal terms (terms with $s$) in $S$ and all non-periodic terms (terms without $s$) in $R$. Similarly, a multiplicative
trend can occur when $\mathcal{F}(t_i) = R \times S \times A(t_i)$. To fit a light curve with multiplicative trend, the idea is to take the logarithm of the fluxes,
$\log\mathcal{F}(t_i) = \log R + \log S + \log A(t_i)$, and then to fit an ARIMA model with an additive trend \citep{3b1355aedd1041f1853e609a410576f3}. This transformation ensures
that the present emission is a weighted geometric mean of the past emissions and fluctuations instead of an arithmetic mean appropriate for the case of an additive trend. This does
not affect the physical interpretation as long as the time-series points are positive, as is the case with blazar light curves. Both a linear additive and an exponential
multiplicative trend were considered while fitting the light curves. Logarithmically transforming the flux (for multiplicative trends) changes the ordinate of the light curve,
which in turn changes the likelihood, and thus the AIC. Thus we cannot directly compare AICs after fitting a log-transformed light curve to the original light curve. To take care
of that, we need to add the Jacobian of the transformation to the AIC of the transformed fit \citep{10.2307/2988185}, which, for the case of a log-transform is $2\sum_i \log\
\mathcal{F}(t_i)$.

Light curves often have long-term memories that can mimic an overall trend. Though a long-term memory (correlation) is difficult to physically explain in the context of the Doppler
boosting understood to dominate blazar emission, the present work explores it nonetheless for the sake of completeness. In the presence of long term memories in the light curve, it
is necessary to use a fractional value of $|d|<1 \in \mathbb{R}$. This was implemented by binomially expanding the $\Delta^d$ operator \citep[][]{ARFIMA1,
2018FrP.....6...80F},
\begin{equation}
  \label{eq3}
  (1-\mathcal{L})^d = \sum_{i=1}^{\infty} \frac{\Gamma(i-d)}{\Gamma(-d)\Gamma(k+1)}\mathcal{L}^i,
\end{equation}
resulting in a Autoregressive (AR) Fractionally Integrated (FI) Moving Average (MA) model, or ARFIMA$(p,|d|<1,q)$. While looking for periodicities, we compared the AIC values of
ARIMA and ARFIMA models with their seasonal counterparts. We also looked for the overall trend (including long-term memory) that best explained the light curve.

The present work used both optical and $\gamma$-ray band data while searching for periodicities. Optical observatories, being ground-based, are subjected to inconveniences such as
daytime transit of the source and bad weather conditions that degrade the observational cadence of the source in the optical band. To make meaningful inferences, it is preferable
to use the better-sampled waveband. However, the said inferences could be extended to the worse sampled band if the emission in both the bands is correlated. We used a Discrete
Correlation Function \citep[DCF,][]{1988ApJ...333..646E} to understand the temporal correlation between the emissions in the two wavebands. If emissions in different wavebands show
a high correlation, then even if the dominant period in one band is less significant, we can probably attribute it to poorer sampling (and not to something intrinsic). The DCF was
calculated by first computing the unbinned DCF
\begin{equation}
  \label{eq7}
  U D C F_{i j}=\frac{\left(\mathcal{F}_\gamma (t_i)-\overline{\mathcal{F}_\gamma}\right)\left(\mathcal{F}_V(t_j)-\overline{\mathcal{F}_V}\right)}{\sqrt{\sigma^{2}({\mathcal{F}_\gamma}) \sigma^{2}({\mathcal{F}_V})}} .
\end{equation}
Here $\mathcal{F}_A(t_i)$ is the $A$ band flux at time $t_i$ and $\overline{\mathcal{F}}_A$ and $\sigma^{2}(\mathcal{F}_A)$ are the corresponding flux mean and variance,
respectively. To obtain the DCF, we binned the UDCF by averaging the points with a delay ($\Delta t_{ij}= t_i - t_j$) in the range $\tau - \Delta \tau/2 \leq \Delta\ t_{ij} \leq
\tau + \Delta \tau/2$. Here $\tau$ is the time lag and $\Delta\tau$ is the bin width. The DCF is then
\begin{equation}
  \label{eq8}
\begin{split}
D C F(\tau) & =\frac{1}{n} \sum_{k=1}^n U D C F_{k}(\tau) \\
& \pm \frac{1}{n-1} \sqrt{\sum_{k=1}^{n}\left(U D C F_{k}-D C F(\tau)\right)^{2}} .
\end{split}
\end{equation}

Since the blazar emissions arise from a non-stationary statistical process, the mean and standard deviation were calculated using only the points that fall within a given time-lag
bin \citep{WhitePeterson1994}. Both the light curves were de-trended by subtracting a linear baseline before the DCF analysis, following \cite{Welsh1999}.

\section{Results}
\label{sec:4}

\begin{figure*}
  \centering
  \subfloat{\includegraphics[width=.53\linewidth]{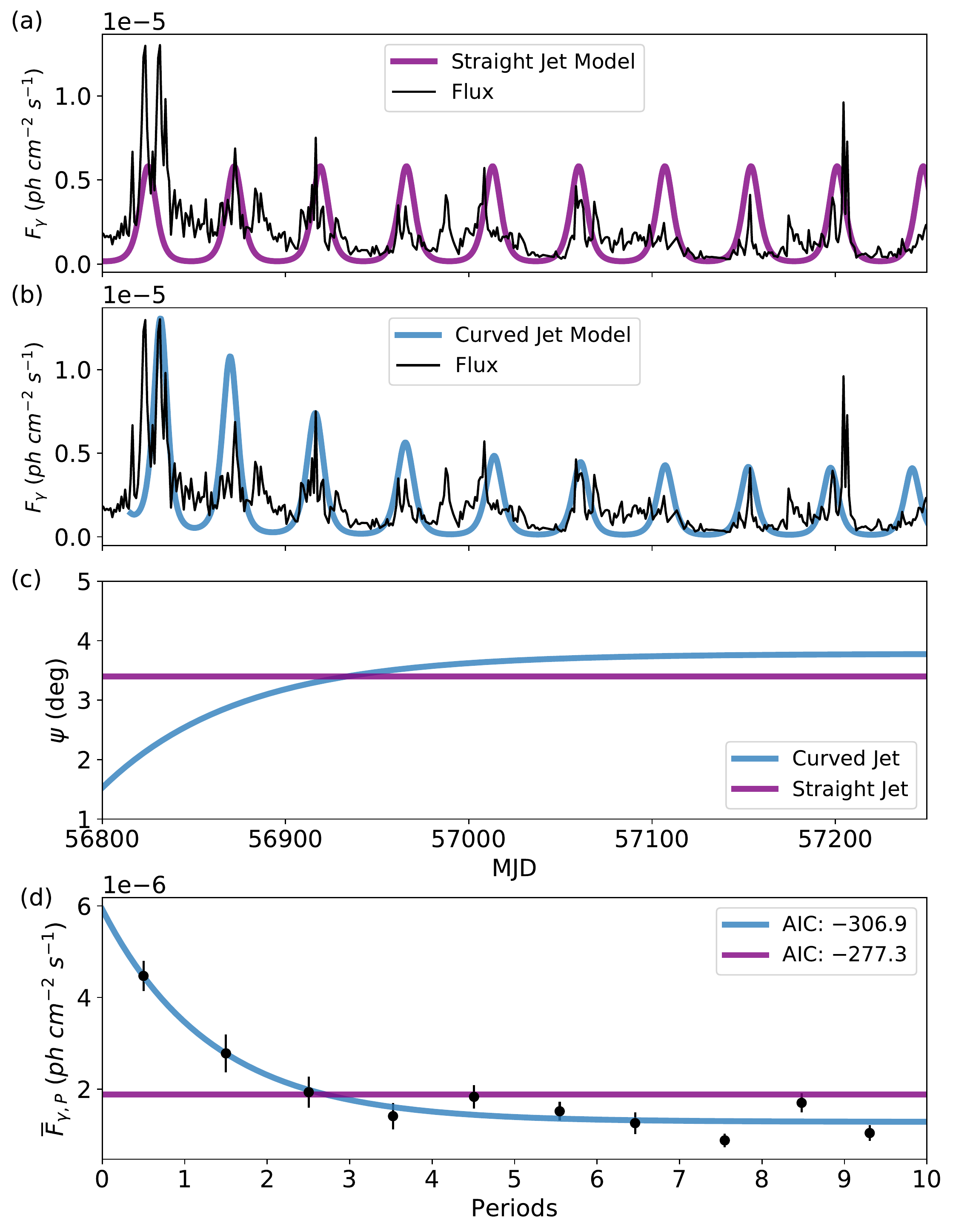}}
  \subfloat{\includegraphics[width=.47\linewidth]{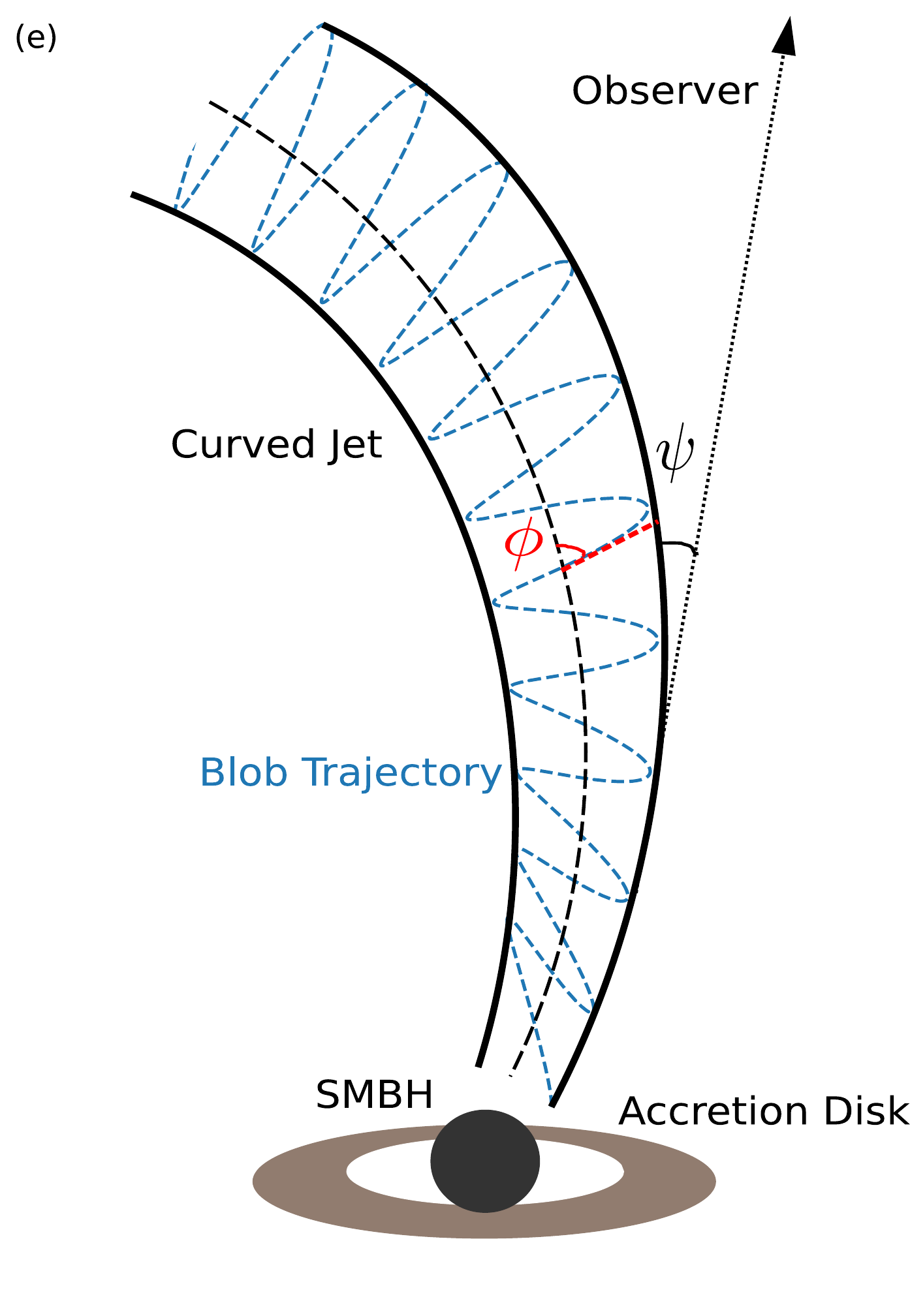}}
  \caption{\label{f7} $\gamma$-ray emission (black) modeled by a blob moving on a helical helical path within \textbf{(a)} a straight jet (purple) or by one moving within
\textbf{(b)} a curved jet \textbf{(cyan)}. \textbf{(c)} Viewing angle of the jet as a function of time (which can also be interpreted as the change in viewing angle as a function of
distance from the base of the jet) for these models. \textbf{(d)} Average flux in each period modeled using straight and curved jets, with the AIC values given in the inset. Lower
AIC values indicate a better model, so the curved jet is a more likely scenario. \textbf{(e)} A cartoon of a blob moving helically inside a curved jet.}
\end{figure*}

A systematic search for quasi-periodicity in both segments A and B of the light-curves in Figure \ref{f1} revealed correlated (albeit transient) QPOs in segment B. No strong
periodicity was observed in segment A and we refrain from further discussing this portion of the data. From the WWZ map of interval B (Figure \ref{f2}a) we observe quasi-periodic
modulation in the $\gamma$-ray flux with the dominant period centered at $47.4^{+0.97}_{-0.51}$ days. The QPO lasts for $\sim 450$ days (MJD 56800 to 57250), demonstrating almost
10 cycles (Figure \ref{f2}b) with a significance $4.1\sigma$ (Figure \ref{f2}c). Periodic modulation is also seen in the optical flux with essentially identical dominant period
centered at $47.3^{+1.08}_{-0.63}$ (Figure \ref{f2}d). The optical QPO lasts for $\sim 250$ days (around 5 cycles) with a significance of $2.4\sigma$ (Figure \ref{f2}f).

Analytically, considering an underlying white noise model, the FAP is $\sim 10^{-5}$ (Figure \ref{f2}c) for the $\gamma$-ray band and $\sim10^{-3}$ for the optical band (Figure
\ref{f2}f). Considering an underlying AR1 process, the significance of the dominant period in both the bands turns out to be $>99\%$ (Figure \ref{f3}a--b). The WWZ significances
were calculated using MC techniques, where the simulated light curves were generated by fitting the PSDs and PDFs of the original light curves using a smooth bending power-law and a
log-normal function respectively, during the suspected QPO (Figure \ref{f9}).

We subdivided the SARIMA$(p,d,q)\times(P,D,Q)_s$ family of models into $10$ different classes ($\mathcal{M}_{1-10}$) based on the overall trend and periodicity, prior to fitting
the light curve, as follows.
\begin{enumerate}
\item $\mathcal{M}_{1-2}$: No overall trend: obtained by setting $A(t)=0$.
\item $\mathcal{M}_{3-4}$: Linear additive trend: obtained by setting $A(t)=mt+c$.
\item $\mathcal{M}_{5-6}$: Multiplicative exponential trend: calculated by first taking $\log \mathcal{F}(t_i)$ (or log-transforming the light curve) and setting $\log A(t)
= mt + c$.
\item $\mathcal{M}_{7-10}$: Long term memories assumed to be present in the light curve: they are obtained by setting $|d|<1\in \mathbb{R}$. For these, no overall trends were
  considered, i.e., $A(t)=0$; specifically, in $\mathcal{M}_{7-8}$ we set $d=0.2$ and $\mathcal{M}_{9-10}$ has $d=0.5$. 
\end{enumerate}
For the above, $\mathcal{M}_{2i}\ i\in\{1...5\}$ consisted of classes with periodic components and $\mathcal{M}_{2i-1}\ i\in\{1...5\}$ consisted of the corresponding non-periodic
control classes, obtained by setting $s=0$. For each class, $\mathcal{M}_i$, we obtained all the models belonging to the class by varying the parameters $\{p,P,d,D,q,Q,s\}\in
\Psi$. We then fitted the \(\gamma\)-ray light curve between MJD 56800 to 57250 using different models to obtain the AIC map. Fitting any light curves to models belonging to the
SARIMA family ideally requires the light curve to be sampled uniformly, although it can account for a moderate amount of unsampled data \citep{2018FrP.....6...80F}. To improve the
sampling rate, the detection threshold was reduced to $3\sigma$ and upper limits wherever present were interpolated. The interpolated value was considered with a probability of
$0.95$ if it is less than the upper limit; otherwise, it was considered with a probability of $0.05$. Following the interpolation, only four datapoints (in 450 days) were absent,
ensuring a sampling rate of $99.1$ per cent.

Figure \ref{f4} shows the AIC values for several models (a subset of $\Psi$) belonging to the model classes $\mathcal{M}_{1-6}$. We observe that the models with an exponential
multiplicative trend ($\mathcal{M}_{5-6}$) give a much better fit to the $\gamma$-ray light curve than do any of the other models considered. The origin of this trend is discussed
in the subsequent section. According to the AIC, the most likely model for the $\gamma$-ray light curve turned out to be SARIMA$(4,1,1)\times(1,0,0)_{47}$, a periodic model with a
period of 47 days, exactly as observed from the LSP and WWZ analyses. While scanning different periods, the AIC plot shows a significant dip at $s=47\text{d}$ (Figure \ref{f4}j)
implying that models including a period of 47 days best describe the light curve. The AIC maps for a subset of fractionally integrated models ($\mathcal{M}_{7-10}$) are given in
Figure \ref{f5}. They were obtained by first differencing the light curve following Eqn. \ref{eq3} (as shown in Figure \ref{f5}e) and subsequently fitting a
SARIMA$(p,0,q)\times(P,0,Q)_s$ model to the data. Even though periodic fractionally integrated models better explain the light curve than do the non-periodic ones, the overall AIC
values for these classes of models are much worse than for $\mathcal{M}_{5-6}$.

The analytical form of the model that fits the $\gamma$-ray light curve the best is
\begin{equation}
  \label{eq9}
  \begin{split}
    \mathcal{F}(t_i) & = e^{-at}\times\mathcal{F}(t_{i-47\text{d}})\times  \mathcal{F}(t_{i-1})\\
    & \times \left\{ \prod_{j=1}^4 \left( \frac{\mathcal{F}(t_{i-j})}{\mathcal{F}(t_{i-j-1})} \right)^{\theta_j} \prod_{j=0}^1 \epsilon(t_{i-j})^{\phi_j}\right\}
    \end{split}
\end{equation}
where $a=3.01\times10^{-3}\ \text{d}^{-1}$ is the coefficient of the multiplicative trend, and $t\equiv (t-56800)\text{d}$. The best fit models from each class and their AIC values
are given in Table \ref{t1}. The residuals of the fits for the best model in each class (except fractional models) as well as the PSF, PSD, and autocorrelation function (ACF) of
the said residuals are given in Figure \ref{f6}. It can be seen that periodic models have lower residual PSD powers at around 47 days period, implying a better fit.

In the present study, we model only the $\gamma$-ray light curve from MJD 56800 to 57250. The optical light curve was not modeled because of its poorer sampling. In fact, optical
observations were not available beyond MJD 57050, due to the daytime transit of the source. However, the $\gamma$-ray and optical emissions seem to be nicely correlated at zero lag
during the portion of segment B where data for both bands are available (Figure \ref{f1x}a). Thus, it seems likely that the lower significance of the optical QPO is due to the
absence of data for the remainder of segment B.

\section{Discussion}
\label{sec:5}
Our analyses from Section \ref{sec:4} demonstrate a $\sim$  $4\sigma$ QPO in the $\gamma$-ray light curve of the blazar 3C 454.3. This periodic flux modulation lasted for 450 days
(MJD 56800 to 57250) with a dominant period of $\sim47$ days. We also observe periodic flux modulation in the optical waveband. Even though the formal significance of the optical
quasi-period is $< 3\sigma$, we still consider this period significant because the optical and $\gamma$-ray QPOs have the same periods and occur simultaneously. It is thus
extremely likely that the periodicity in both the wave-bands arises from closely related intrinsic physical processes and the significance of the optical QPO is lower mostly
because of the poorer sampling in that band.

Along with the QPO, the modeling of the light curve using a SARIMA process also identified an exponentially decaying multiplicative trend during the QPO period which also needs to
be explained. A multiplicative trend in any time series is a signature of its variance depending on its mean. Visually, we observe this in the $\gamma$-ray light curve where the
variance decreases with a decrease in flux during the QPO period. In the context of blazars, a multiplicative trend naturally can have its origin in the Doppler boosting of the
radiation. The emission in the rest frame and Doppler boosted frame are related by $\mathcal{F}(\nu)\propto\delta^{3}\mathcal{F}_{\nu'}$ (primed coordinates are in the rest frame
of the blob) thus any intrinsic variation in the rest frame $\Delta \mathcal{F}_{\nu'}$ is amplified by a large factor proportional to $\delta^{3}$ \citep{UrryPadovani1995}. Hence
any change in $\delta$ could manifest as a multiplicative trend in the blazar light curve.

Ignoring the trend, several different physical models might explain the origin of periodicity or quasi-periodicity in a blazar emission. One possible explanation could be a binary
supermassive black hole (SMBH) AGN system \citep{2015ApJ...813L..41A, 2016AJ....151...54S, 2016ApJ...820...20S} where the orbital period provides the flux modulation. The total
mass of the binary SMBH system in those cases is $\sim 10^8 M_{\odot}$, while 3C 454.3 is almost certainly more massive \citep{2002ApJ...579..530W}. So the secondary BH's orbital
period would be of the scale of a few years considering the distance between them to be of the order of milli-parsec. Explicitly, periodicity can be induced by the secondary BH by
piercing the accretion disc of the primary BH during the orbit \citep{2008Natur.452..851V} and for the blazar OJ 287, where the mass is greater, the period is $\sim 12$ y. Also, a
binary SMBH model cannot naturally explain the fading of the oscillation in 3C 454.3 after around $500$ days. 

A second possibility involves an accretion disc hotspot orbiting close
to the innermost stable circular orbit of the SMBH \citep[e.g.][]{2019MNRAS.484.5785G}. Hotspot emissions are quasi-thermal and could directly explain the optical flux modulation.
The optical emission could produce variations in the seed photon field for external Compton interactions which gives rise to the flux modulation in the $\gamma$-ray light curve
\citep{2017MNRAS.472..788G}. However, the expected period would be much shorter than 47 days for any reasonable SMBH mass and spin. Further, it would not be the same as the
$\gamma$-ray period, which would be Doppler boosted. While jet precession models could give rise to QPOs in the light curve of a blazar, the expected period is $> 1$ year
\citep{2004ApJ...615L...5R} which does not agree with the present observations.

\begin{figure}
\centering
\includegraphics[width=.9\linewidth]{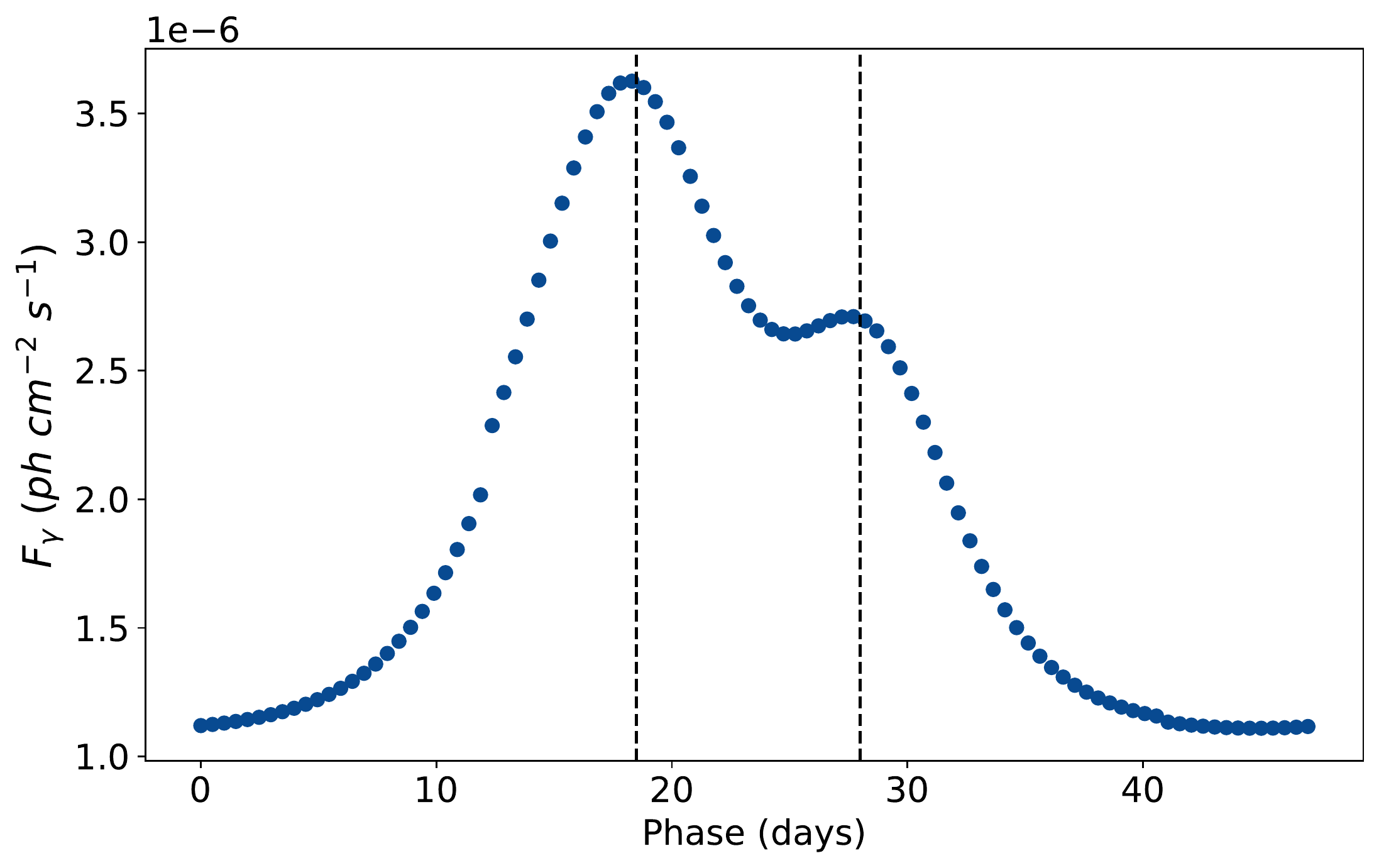}
\caption{\label{f8}
Folded light curve obtained by phasing the modeled emission (blue curve in Figure \ref{f7}b) at a period of 47 days. The two peaks are separated by $\sim 10$
days; we note that this is very similar to the observations in Figure\ \ref{f3}c.}
\end{figure}

We find that the most likely scenario for the observed QPO is for it to come from a region of enhanced emission, or blob, moving helically within the jet \citep[see][for a general
relativistic treatment of the scenario]{2015ApJ...805...91M}. The Doppler factor is related to the viewing angle of the emission region and the motion of the blob changes the
Doppler boosting and can result in a significant change in the observed flux. For a blob moving helically, the changing viewing angle of the blob with respect to our line of sight,
$\theta_{obs}(t)$, is given by \citep{2017MNRAS.465..161S}
\begin{equation}
  \label{eq10}
  \cos\theta_{obs}(t) = \sin \phi \ \sin \psi \cos (2\pi t/P_{obs}) + \cos \phi \ \cos\psi ,
\end{equation}
where $P_{obs}$ is the observed period, $\phi$ is the pitch angle of the helix and $\psi$ is the angle of the axis of the jet with respect to our line of sight (Figure\ \ref{f7}e).
The Doppler factor dependence on viewing angle is $\delta=1/\Gamma(1-\beta \cos \theta_{obs})$, where $\Gamma = 1/(1-\beta^{-2})^{1/2}$ is the bulk Lorentz factor and
$\beta=v_{jet}/c$. Substituting the value of $\cos \theta_{obs}(t)$ in the expression for $\delta$ (since $\mathcal{F}_{\nu}\propto \delta^3$), we get,
\begin{equation}
  \label{eq11}
  \mathcal{F}_{\nu} \propto \frac{\mathcal{F}'_{\nu'}}{\Gamma^3(1+S)^3} \left(1-\frac{\beta C}{1+S}\cos(2\pi t/P_{obs})\right)^{-3}
\end{equation}
where $\mathcal{F}'_{\nu'}$ is the rest frame emission, $P_{obs}$ is the observed period, $C \equiv \cos\phi \ \cos\psi$, and $S \equiv \sin\phi \ \sin\psi$. The overall
multiplicative trend that was plausibly attributed above to changes in the Doppler factor could certainly result from a spatial curvature in the jet that manifests as a change in
viewing angle (and by extension Doppler factor) with time as the blob moves downstream. Thus we modeled the boosted emission in the observed frame by considering both a straight
jet (Figure\ \ref{f7}a) and a curved jet (Figure\ \ref{f7}b) with the viewing angle, $\psi \equiv \psi(t)$, to the jet (at the position of the blob as a function of time) given in
Figure\ \ref{f7}c. The functional form of $\psi(t)$ was obtained by inverting Eqn.\ \ref{eq11} for the observed multiplicative trend factor $a$ in Eqn.\ \ref{eq9}.

Judging by the AIC, both the observed flux and the flux per period is better modeled by a curved jet (Figure\ \ref{f7}d). Due to Doppler boosting, the periods in the observed and
rest frame of the blob are related by $P_{obs} = P_{rest}(1-\beta \cos \psi(t)\cos \phi )$, implying either $P_{obs}$ or $P_{rest}$ is a function of time if the other one is
constant and the jet is curved. We make the more physical assumption that the period in the rest frame of the blob is constant, so $P_{rest} = P_{obs}/(1-\beta \cos \langle \psi(t)
\rangle \cos \phi )\approx 27.6$ years, which translates to a change in the observed period from 40 to 48 days. Since nearly 70 per cent of the observed stretch is dominated by
near-constant viewing angles, the resultant dominant quasi-period is closer to 47 days.

Light curves folded with a period of 47 days (Figure \ref{f3}) were fitted using a Lorentzian approximation of Equation \ref{eq11}. Both $\gamma$-rays and V-band required two
Lorentzian components, $\mathcal{L}_1$ and $\mathcal{L}_2$, to properly fit the folded light curve. The peak positions of the two Lorentzian components are the same in both the
$\gamma$- and V-bands, but the relative amplitudes of the two components are different. This is similar to recent observations of the folded \(\gamma\)-ray lightcurve of the blazar
PKS 2247$-$131, where the two features were explained as a signature of two discrete emission regions, perhaps corresponding to forward and backward shocks
\citep{2018NatCo...9.4599Z}.

The present model with the variable period of oscillation (due to jet curvature) provides another plausible explanation for the two Lorentzian components in the folded light curve.
Here $\mathcal{L}_1$, is contributed by aggregating the flux from most of the cycles (3 onwards), whereas the component $\mathcal{L}_2$ is due to fluxes from cycles 1--3, where the
greater differences in period translates to a difference in phase in the folded $\gamma$-ray light curve. Since the observed fluxes in the first three cycles are significantly
higher than the rest, their contribution does not wash out, even after averaging over 10 periods (see Figure \ref{f8}). However, the V-band light curve was observed for fewer cycles
($< 5$), so the folded light curve is dominated by fluxes from cycles 1--3. Hence $\mathcal{L}_2$ would have a higher relative amplitude for the V-band, as observed. We also note
that the optical electric vector polarization angle (EVPA) varied substantially, from $0^{\circ}$ to $180^{\circ}$ during this period (see Figure \ref{f1x}b). Such variations are
another signature of blobs moving helically in the jet \citep{2008Natur.452..966M}, but in the present case, the polarimetric sampling is not good enough to adequately define the
evolution of the EVPA with time.

Our analysis strongly suggests that the observed modulation of the flux from 3C 454.3 over more than a year is due to an enhanced emission region moving helically within a curved
jet or conceivably a curved-helical jet itself. But since blazar emission mechanisms are not well understood, it is possible that the observed flux modulation is due to some other
effect (intrinsic or otherwise) or a combination of such effects. Considering the model to be correct with parameter values $\phi\approx 2^{\circ}$ \citep{2017MNRAS.465..161S} and
$\Gamma=15$ \citep{Abdoetal2011}, we can best model the viewing angle as changing from $2.6^{\circ}$ at MJD 56850 to $3.6^{\circ}$ over three QPO cycles (after the first, partial
one, which corresponds to an initial $\psi$(MJD 56800) = 1.6$^{\circ}$ (Figure \ref{f7}c). The distance travelled by the blob in one period is $D_{1P} = c\beta P_{rest}\cos\phi
\approx 8.4$ pc. Thus the viewing angle changes by $1^{\circ}$ over a distance of $3D_{1P}$, resulting in the jet curvature to be $\approx 0.05^{\circ}$ pc$^{-1}$.

\section*{Acknowledgements}
\label{Sec:7}

We thank the anonymous referee for suggestions that improved the presentation of our results. This research has used data, software, and web tools of the High Energy Astrophysics
Science Archive Research Center (HEASARC), maintained by NASA's Goddard Space Flight Center. Data from the Steward Observatory spectropolarimetric monitoring project were used.
This program is supported by Fermi Guest Investigator grants NNX08AW56G, NNX09AU10G, NNX12AO93G, and NNX15AU81G. This paper has made use of up-to-date SMARTS optical/near-infrared
light curves that are available at \href{www.astro.yale.edu/smarts/glast/home.php}{www.astro.yale.edu/smarts/glast/home.php}. AS and VRC acknowledge support of the Department of
Atomic Energy, Government of India, under project no. 12-R\&D-TFR-5.02-0200.

\section*{Data Availability}

\textit{Fermi}-LAT: \href{https://fermi.gsfc.nasa.gov/ssc/data/access/}{https://fermi.gsfc.nasa.gov/ssc/data/access/}\\
SMARTS: \href{www.astro.yale.edu/smarts/glast/home.php}{www.astro.yale.edu/smarts/glast/home.php}\\
SPOL: \href{http://james.as.arizona.edu/~psmith/Fermi/DATA/Objects/targets.html}{http://james.as.arizona.edu/~psmith/Fermi/DATA/Objects/}\\

\bibliography{QPO}
\bibliographystyle{mnras}
\end{document}